\documentclass[useAMS]{mn2e}

\usepackage{graphicx}
\usepackage{graphics}
\usepackage{natbib}
\usepackage{subfigure}
\usepackage{url}

\newcommand{\aap}{A\&A}

\newcommand{\aapr}{A\&AR}
\newcommand{\aj}{AJ}
\newcommand{\apj}{ApJ}
\newcommand{\apjl}{ApJL}

\newcommand{\araa}{ARA\&A}
\newcommand{\mnras}{MNRAS}

\newcommand{\pasp}{PASP}

\newcommand{\nb}{Nuclear Bulge}

\newcommand{\tmass}{{\it 2MASS}}

\newcommand{\ukidss}{{\it UKIDSS}}
\newcommand{\vista}{{\it VISTA}}

\newcommand{\vlt}{VLT}

\newcommand{\msol}{M_{\odot}}

\newcommand{\ks}{\ensuremath{K_{\rm S}}}

\newcommand{\jh}{J\!-\!H}
\newcommand{\jk}{J\!-\!K_{\rm S}}
\newcommand{\hk}{H\!-\!K_{\rm S}}
\newcommand{\nir}{near-infrared}

\newcommand{\ltsim}{\mbox
{{\raisebox{-0.4ex}{$\stackrel{<}{{\scriptstyle\sim}}$}}}}

\title[nIR extinction towards the Nuclear Bulge]{The complex, variable
near-Infrared extinction towards the Nuclear Bulge}

\author[A. J. Gosling et al.]{Andrew J. Gosling,$^{1,2}$\thanks{e-mail:
andrew.gosling@oulu.fi} Reba M. Bandyopadhyay,$^{3}$ and Katherine
M. Blundell$^{1}$\\ 
$^{1}$Department of Astrophysics, University of Oxford, Keble Road,
Oxford, OX1 3RH, UK \\
$^{2}$Astronomy Division, Department of Physical Sciences, P.O. Box 3000, 90014 University of Oulu, Finland \\
$^{3}$Department of Astronomy, University of Florida, Gainesville, FL
32611, USA}

\begin{document}

\date{\today}

\maketitle

\label{firstpage}

\begin{abstract}

Using deep $J$, $H$ and \ks-band observations, we have studied the
\nir\ extinction of the \nb\ and we find significant, complex variations
on small physical scales. We have applied a new variable \nir\ colour
excess method, V-NICE, to measure the extinction; this method allows for
variation in both the extinction law parameter $\alpha$ and the degree
of absolute extinction on very small physical scales.  We see
significant variation in both these parameters on scales of
$5\arcsec$.  In our observed fields, representing a random sample of
sight lines to the \nb, we measure $\alpha$ to be $2.64 \pm 0.52$, 
compared to the canonical ``universal'' value of 2.  Our measured levels of
$A_{\ks}$ are similar to previously measured results ($1 \leq A_{\ks}
\leq 4.5$); however, the steeper extinction law results in higher
values for $A_J$ ($4.5 \leq A_J \leq 10$) and $A_H$ ($1.5 \leq A_H
\leq 6.5$).  Only when the extinction law is allowed to vary on the
smallest scales can we recover self-consistent measures of the
absolute extinction at each wavelength, allowing accurate reddening
corrections for field star photometry in the \nb.  The steeper
extinction law slope also suggests that previous conversions of
\nir\ extinction to $A_V$ may need to be reconsidered.  Finally, we
find that the measured values of extinction are significantly dependent 
on the filter transmission functions of the instrument used to obtain the
data.  This effect must be taken into account when combining or 
comparing data from different instruments.

\end{abstract}


\section{Introduction}

Study of the structure and stellar content of the \nb\ is extremely
difficult as it is one of the most highly obscured regions of the
Galaxy. The Nuclear Bulge is estimated to contain a narrow layer of
molecular hydrogen throughout its extent with scale height $h \sim
30-50\,{\rm pc}$ and a total mass of $M \sim 10^8\,\msol$
\citep{mezg96,rodr05}. However, this accounts for only half of the
extinction between the observer and the Galactic Centre as it is
estimated that there is an equal amount of material between the Sun
and the \nb\ as there is within the \nb\ in the direction of the
Galactic Centre \citep{gord03}.

If this material were distributed homogeneously over the sky, it would
result in visual extinction of $\geq 100\,{\rm mag}$. Numerous authors
have measured the extinction in the visual and near-infrared and found
the level of extinction to be $A_V \sim 30\,{\rm mag}$ \citep[][and
others]{riek85, catc90, blum96, dutr02, dutr03}. The reason for this
discrepancy is that only $\sim 10\%$ of the interstellar medium is
distributed homogeneously throughout the \nb\ and the remaining $\sim
90\%$ is contained within dense clouds filling only a few percent of
the \nb\ \citep{catc90}.

\subsection{The Extinction Law}
\label{s:intextlaw}

The relationship between extinction and wavelength is governed by the
composition of the material causing the extinction; in most cases,
this is a diffuse mixture of gas and dust \citep[small grains and
molecules;][]{fitz04}, and is known as the ``extinction law''. In the
visual wavelength regime, the extinction law is generally expressed as
the ratio of absolute to relative extinction $R_V = A_V/E_{B-V}$. In
the near-infrared wavelength range the extinction law is described by
a power-law relationship, $A_{\lambda} \propto \lambda^{-\alpha}$, and
can be expressed either as a ratio of relative extinctions
$E_{J-H}/E_{H-K}$ or, as the slope of the power-law $\alpha$
\citep[][and others]{riek85,card89,math90}.

The original, and generally applied, extinction law for the diffuse
interstellar medium (i.e.: not dense, known molecular clouds) was
calculated in the 1980s using relatively primitive (by today's
standards) infrared detectors and using a small sample of individually
selected stars whose spectral types were well known
\citep{sava79,riek85, card89, math90, catc90,whit93}.  Based on the
small sample of sources for which they had spectral observations, they
expressed the near-infrared extinction law as ``universal'', a notion
that has been adopted in almost all studies since. There have been
many values given for the extinction law, with values of $E_{J-H}/E_{H-K}$
ranging from 1.28 to 2.09 \citep{davi86,jone80}; however, this variation
has been attributed to the different magnitude systems being used
\citep[see][for an example of photometric systems affecting the
extinction law determination]{keny98}. Up to this point the possibility
that the underlying extinction law could vary on small scales had not
been considered extensively.

However, in the last decade, as infrared detector technology has
advanced, it has been possible to obtain much higher-resolution,
deeper surveys of the sky in the infrared band (\tmass, {\it DENIS},
\ukidss, \vista). Based on these observations, it has become evident
that the extinction law is not ``universal'' as previously thought,
but is in fact highly variable from point-to-point
\citep{keny98,udal03,mess05, nish06, fitz07, froe07}. This variation
is most evident towards the Bulge and in denser regions such as
molecular clouds \citep{drai03}. \citet{nish06} have shown there is
variation in the extinction law within their observations of the
\nb. They measure $E_{J-H}/E_{H-K} = 1.72 \pm 0.04$, equivalent to an
$\alpha = 1.99 \pm 0.02$ for the outer regions of the Bulge, but they
find a spatial variation with values of $\alpha$ of 1.96, 1.97, 2.09
and 1.91 for the North-East, South-East, North-West and South-West
regions respectively.  \citet{froe06} demonstrated how the assumption
of a power law extinction-wavelength relationship can be manipulated
to provide equations to measure that relationship given photometry in
three wavebands; this is the method we will utilize in this
paper. They applied this method to the Galactic anti-centre using data
from \tmass\ \citep{froe07} and found a large spread in the measured
extinction law parameter $\alpha$. However, when calculating the
absolute extinction, they took an average of the spread in extinction
law parameter to convert from colour excess to extinction. Most
recently, \citet{nish08} showed in observations of the outer regions
of the Galactic Bulge that the conversion factor between near-infrared
and visual extinction was much higher ($A_V/A_K \sim$16, compared
to the canonical value of $\sim$9) than previous studies implied,
indicative of a steeper extinction law slope (a higher value of
$\alpha$).

\subsection{Previous Extinction Surveys}
\label{s:intprevext}

The first large scale extinction mapping of the \nb\ in the
near-infrared was undertaken by \citet{catc90} using $H-K$ and
$K\,{\rm mag}$ from observations on the 1.9\,m telescope at SAAO
Sutherland. They measured the modal colours and magnitudes from
histograms of all the stars in in $400\arcsec \times 400\arcsec$
cells, and compared these values to the giant branch of 47 Tuc to
measure $5 \leq A_V \leq 35$. They observed a general trend of
increasing extinction levels towards zero latitude in the plane, and
also made a ``hand-drawn'' map of the distribution of dark clouds
obscuring all background sources in their observations. Additionally,
in a number of the histograms of colours and magnitudes of the stars,
they observed a double peak in the distributions indicating more than
one source of extinction along the line of sight.

\citet{schu99} used data from the {\it DENIS} survey of the Bulge to
form an extinction map with a resolution of $1\arcmin$. They used a
two-to-one over-sampling, so for their resolution they were sampling
stars within a $2\arcmin$ radius. They used the extinction law of
\citet{glas99} to convert their measured colours to absolute
extinction. They measured $A_V$ as high as 30, although in many
regions this was only a lower limit as in high extinction regions the
majority of sources were obscured, especially in $J$. They also
reported the presence of a clumpy and filamentary distribution to the
areas of high extinction, especially close to the Plane.

The last large-scale extinction mapping of the Bulge before this work
was performed by \citet{dutr03} over $10\deg \times 10\deg$ using data
from \tmass. They sampled $4\arcmin \times 4\arcmin$ regions and
measured extinction ranging from $A_{\ks}$ = 0.005 at edges of the
region to $A_{\ks}$ = 3.2 close to Galactic Centre. As with the
\citet{catc90} map, they observed a general reduction in the level of
the extinction away from the plane, and also a filamentary structure
to the regions of highest extinction.

\section{Extinction Theory}
\label{s:theory}

As discussed in Section 1, the extinction in the \nir\ is
generally accepted to have a power law relationship to wavelength: 

\begin{equation} \label{e:ext}
A_{\lambda} = C \times \lambda^{-\alpha}. 
\end{equation}

where $C$ is a dimensionless constant.  We use a similar method for
measuring the extinction law and the value of absolute extinction as
outlined and discussed in \citet{froe06}.  Colour excess, $\left< E (
\lambda_1 - \lambda_2 ) \right>$, for an object is the difference
between the observed colour for that object, and its intrinsic colour
if there were no extinction:

\[
\left< E ( \lambda_1 - \lambda_2 ) \right> = \left( m^{\rm obs}_1 - m^{\rm obs}_2
\right) - \left( m^{\rm int}_1 - m^{\rm int}_2 \right).
\]

As the observed magnitudes are a function of the intrinsic magnitude
and the extinction at that wavelength, $m^{\rm obs} = m^{\rm int} +
A_{\lambda}$, the colour excess becomes

\[
\left< E ( \lambda_1 - \lambda_2 ) \right> = A_{\lambda_1} - A_{\lambda_2} 
\]

\noindent
which allows the colour excess to be expressed as a function of two
wavelengths and the dimensionless constant $C$

\begin{equation} \label{e:ext1}
\left< E ( \lambda_1 - \lambda_2 ) \right> = C \times ( \lambda_1^{-\alpha} - \lambda_2^{-\alpha} )
\end{equation}

Using Eq. \ref{e:ext1} for two different colour excesses with a common
wavelength, it is then easy to express the ratio of the two different
colour excesses as a function of only wavelength and the extinction
law slope $\alpha$.

\begin{equation} \label{e:extlaw}
\frac{\left< E ( \lambda_1 - \lambda_2 ) \right>}{\left< E ( \lambda_2 -
\lambda_3 ) \right>} = \frac{\left( \frac { \lambda_{2} }{ \lambda_{1} }
\right)^{\alpha} - 1}{1 - \left( \frac { \lambda_{2} }{ \lambda_{3} }
\right)^{\alpha} }
\end{equation}

By comparing the measured \nir\ colour excess variation across images,
the extinction law can then be calculated at each spatial position,
the value of $C$ can then be derived for each location using
Eq. \ref{e:ext1} and thus the absolute value of extinction for a
certain wavelength can be calculated using Eq. \ref{e:ext}.  This is a
more detailed (in that it considers the extinction law $\alpha$
variation as well as absolute extinction) and variable version of the
NICE and NICER extinction calculation methods of \citet{lomb01} which
we term V-NICE.

\section{Data}
\label{s:data}

This work uses data from deep imaging of the \nb\ of our Galaxy,
within a region $1.6^{\circ} \times 0.8^{\circ}$ in $(l,b)$ centred on
Sgr A*. We observed 26 locations within the \nb\ using the ISAAC
camera on the VLT \citep{band05}. The 26 fields are located in areas
away from known regions of star formation or other known
structures within the \nb. The ISAAC instrument has a $2.5 \times 2.5$
arcmin$^2$ field of view with a pixel resolution of 0.1484\arcsec. The
observations were obtained on nights with $\ltsim$0.6\arcsec\
seeing. Each field was observed for a total of 360\,s in each filter,
$J$, $H$ and $\ks$. A random 20\arcsec\ offset was used between
integrations using the standard ISAAC jitter templates to allow
self-flattening of the images. The total 6 minutes integration per
filter gives limiting magnitudes of $J$=23 (S/N=5), $H$=21, and
$\ks$=20 (S/N=10).

Initial reduction (flat-fielding, removal of bad pixels, and sky
subtraction) was performed with the ESO/ISAAC pipeline reduction
software. These image products were then astrometrically locked to the
Two Micron All-Sky Survey (\tmass). Source positions and magnitudes
were extracted using \texttt{SExtractor} \citep[version 2.3.2;
  see][for further details]{band05}. We used sources that were common
to both the \tmass\ catalogue and our own observations to match the
photometry of our sources to that of \tmass. This linear conversion to
the \tmass\ photometric system corrected for slight variations in the
photometric calibration supplied by the ESO/ISAAC pipeline and allows
comparison of this work to the previous extinction mapping of this
region which was based on \tmass\ data \citep{dutr03,froe07}. The
photometric and astrometric data for the sources were combined into a
series of catalogues, one for each of our 26 VLT fields.

\section{Calculating the Extinction}
\label{s:calcext}

We generated a grid of positions across the fields at each of which
the extinction is calculated. The grid has a $5\arcsec$ separation
between points and the grid is defined around the centre of each
field. At each grid point, all stars within $10\arcsec$ and 
$20\arcsec$ are extracted from the source catalogues and the median
value of the stars' colours (magnitude of the star in a particular
filter e.g.: $J$) and colour-index (difference between the
brightness of a star in two colours e.g.: $\hk$) were calculated for
that grid point. The median is used as it reduces the influence of
foreground stars on the values calculated \citep[the effect of the use of
the median instead of the mean is discussed in][]{froe07}.

\begin{figure}
  \begin{center}
    \includegraphics[trim=10mm 10mm 35mm 15mm, clip, width=\columnwidth]{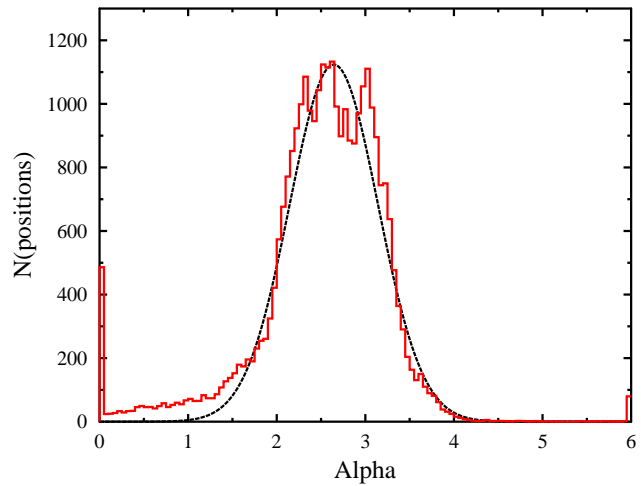}
    \caption[Histogram of extinction law, $\alpha$, calculated from
      the measured colour excess ratios]{Histograms of the value of
      the extinction law parameter $\alpha$, calculated from the
      measured colour excess ratios using Eq. \ref{e:extlaw} (red).
      The best fit single gaussian to the distribution has $\mu =
      2.64$ with $\sigma = 0.52$ is overlaid on the data (black). }
    \label{f:alfhisto}
  \end{center}
\end{figure}

Median colour excess is then obtained by subtracting a value for the
average intrinsic colours for the stars. We used the average of the
colours of G, K and M giant stars as taken from \citet[][and
  references therein]{alle00} to obtain colour excesses for the
\nb\ stars.  We use G, K and M type stars as these are the most common
stellar types in the \nb.  We also set a threshold limit of 5 stars
within the $10\arcsec$ sampling box as a minimum for the number of
stars required for the scripts to calculate a specific extinction. If
there are fewer than this number in the sample region, the value for
that point for both the $10\arcsec$ and $20\arcsec$ sampling is set to
the average value measured for the whole field.

The colour excess ratios for each grid point were then used to
identify the value of the extinction law parameter $\alpha$ by
comparison to tables of values generated using equation
\ref{e:extlaw}. We then used this value of $\alpha$ to calculate the
value of $C$ at that position using equation \ref{e:ext1}, and then
the absolute extinction in $J$, $H$ and \ks-bands was calculated using
equation \ref{e:ext}.

\section{The Extinction}
\label{s:ext}

\begin{figure*}
  \begin{center}
    \includegraphics[trim=10mm 10mm 35mm 15mm, clip, width=0.33\textwidth]{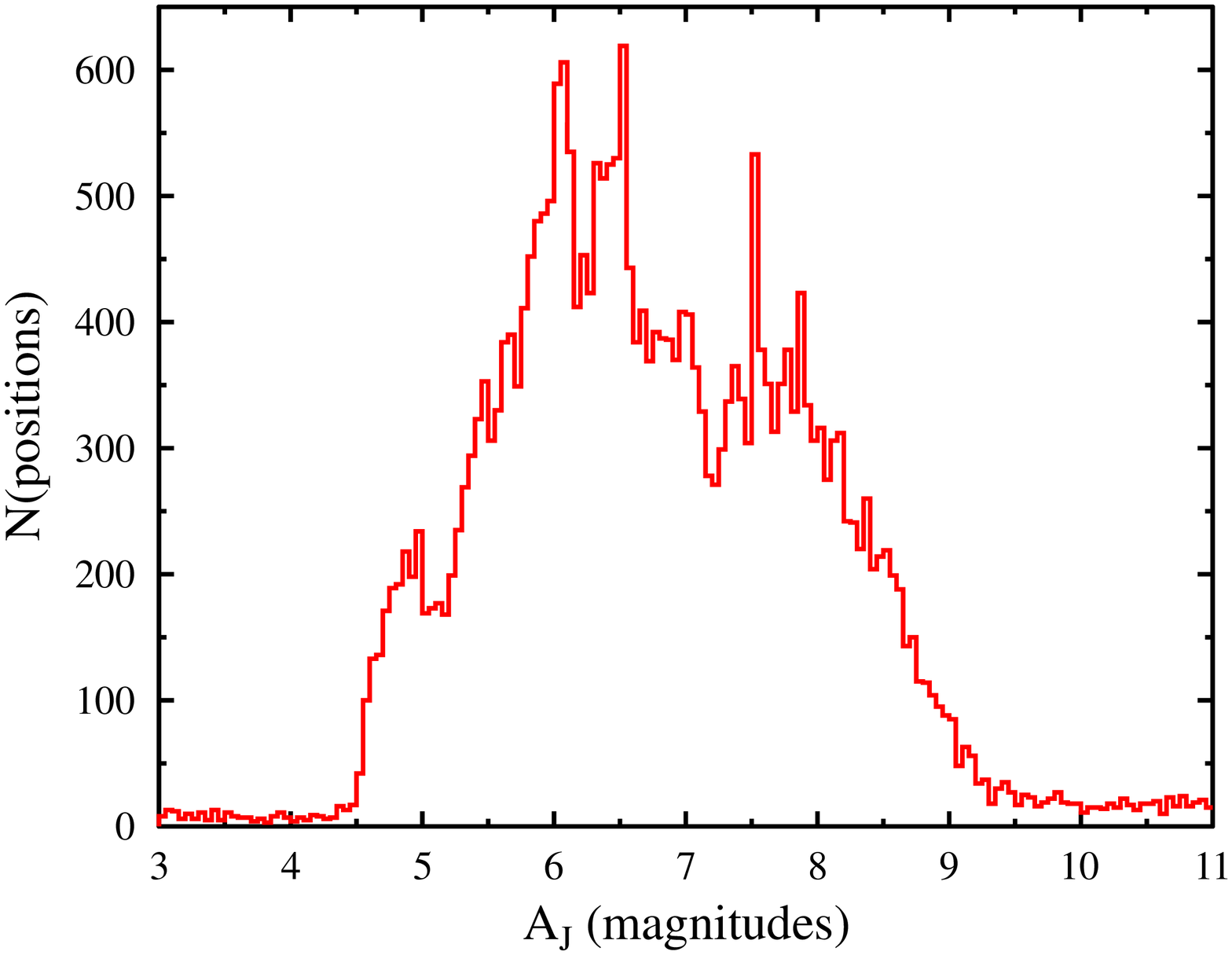} %
    \includegraphics[trim=10mm 10mm 35mm 15mm, clip, width=0.33\textwidth]{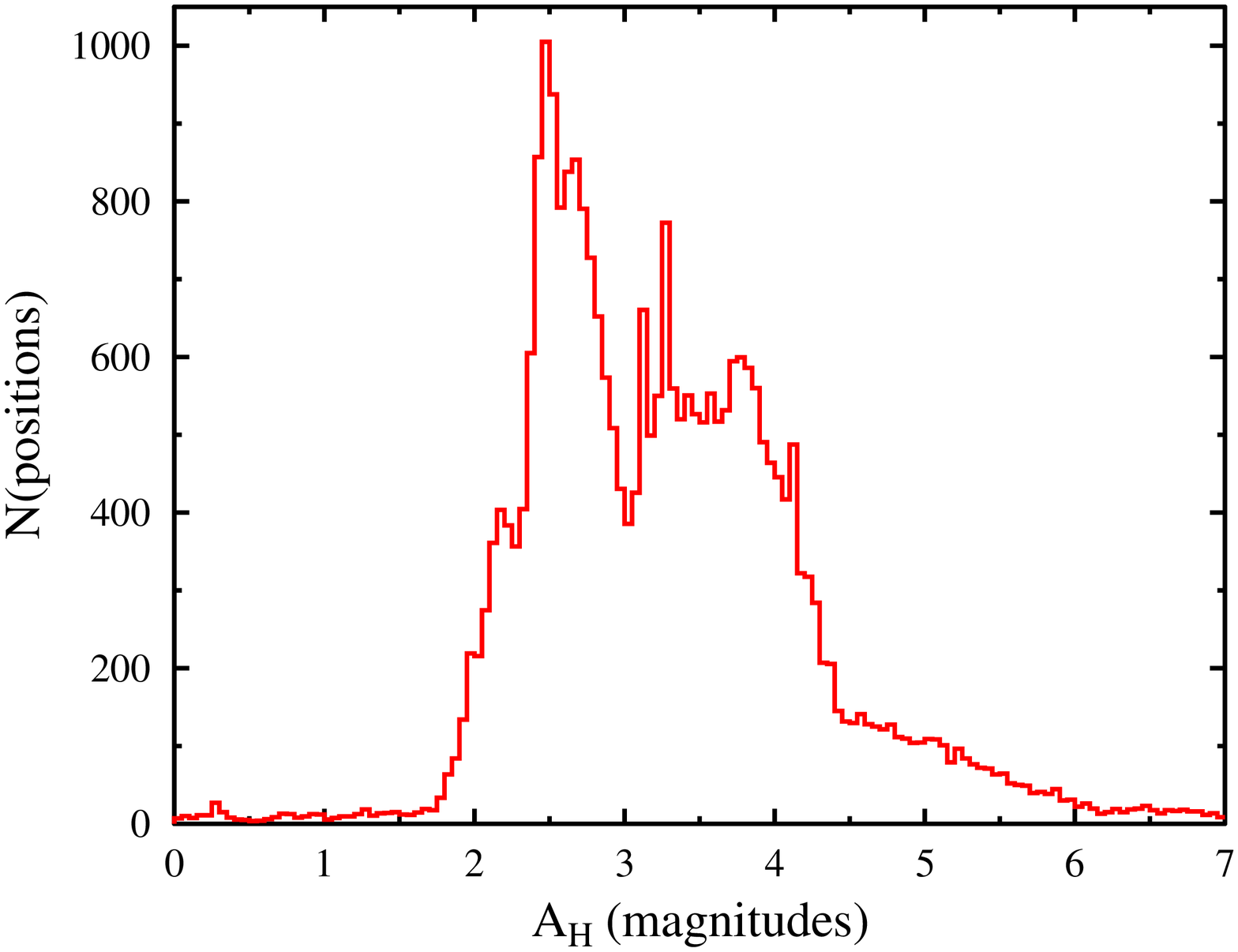} %
    \includegraphics[trim=10mm 10mm 35mm 15mm, clip, width=0.33\textwidth]{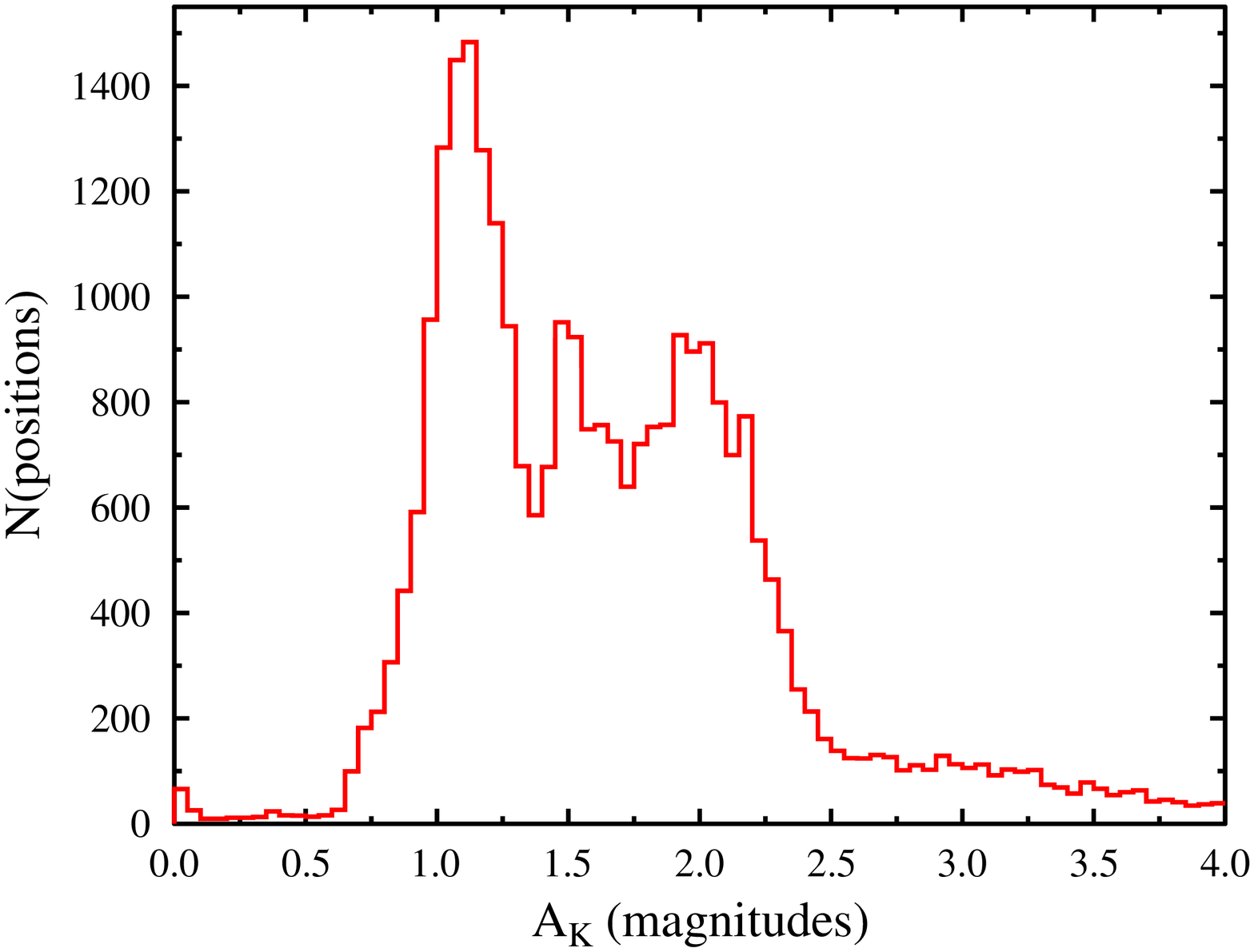}\\ %
    \caption[Histograms of absolute extinction in
      magnitudes]{Histograms of the values of absolute extinction
      calculated for all grid points. Each histogram represents the
      absolute extinction of a specific wavelength; left is $A_J$,
      centre is $A_H$ and right is $A_{\ks}$. 
    }
    \label{f:exthisto}
  \end{center}
\end{figure*}

\subsection{Measured Extinction Law}
\label{s:extlaw}

Figure \ref{f:alfhisto} shows a histogram of the values of $\alpha$
measured for all the fields using the colour excess ratio method in
Eq. \ref{e:extlaw} with a single gaussian fit to the distribution.

The distribution of $\alpha$ is well fit by a single gaussian with
$\mu = 2.64$ and $\sigma = 0.52$. Errors in the intrinsic measured
magnitudes of the stars are the largest source of error in the
measurement of the colour excesses from which each value of $\alpha$
is obtained, but the large sample ($> 30,000$ positions) results in a
vanishingly small error in the mean and standard deviation as measured
from the histograms. The binning for the histograms is 0.05 in values
of $\alpha$; the error in the fit of the gaussian to the histogram is
comparable to one histogram bin.

\subsection{Measured Extinction}
\label{s:measext}

Using the value of $\alpha$ measured at each point in the fields, we
calculated the absolute extinction from the colour excesses using
equation \ref{e:ext1} to obtain a value for $C$ and equation \ref{e:ext} to
calculate the extinction for each wavelength.  Figure \ref{f:exthisto}
shows histograms of the values of absolute extinction for each
wavelength, calculated at all grid points in the 26 VLT fields.

All three distributions of absolute extinction have common
features. The values of extinction are generally confined to a single
range of values, with almost no cells having values outside this
range.  Within this range, there is a small, low-extinction component,
the main component of the extinction is then one large peak at an
absolute extinction a little higher than the low extinction
component. Beyond this initial peak there is a plateau extending to
higher levels of extinction before a relatively sharp cut-off. At the
cut-off there is then a smaller component that forms a tail extending
to very high levels of extinction. That these features are common to
all three wavelengths suggests that the different wavelengths are
showing different representations of a common distribution of
extincting material and not different materials acting at different
wavelengths.

\begin{figure}
  \begin{center}
    \includegraphics[trim=10mm 10mm 35mm 15mm, clip, width=\columnwidth]{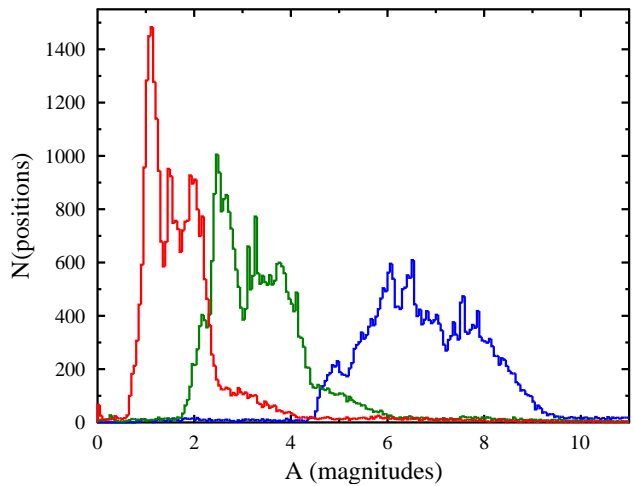}
    \caption[Histogram of extinction in all three bands]{Histogram of
      the amount of extinction in magnitudes at each position, in bins
      of $0.05\,{\rm mag}$. The blue curve (right) is the $A_J$
      distribution, green (middle) is the $A_H$ distribution and red
      (left) is the $A_{\ks}$ distribution.}
    \label{f:allext}
  \end{center}
\end{figure}

The majority of the $A_J$ extinction is in the range $4.5 \leq A_J
\leq 9$, consisting of three peaks.  The low-extinction component of
the distribution peaks at $A_J = 4.8$, the main component of the
distribution peaks around $A_J = 6.1$ and the higher extinction
plateau region is centred around $A_J = 7.8$.

The $A_H$ extinction distribution is in the range $1.5 \leq A_H \leq
6.5$ with a proportionally smaller low extinction component and larger
high extinction component.  The low extinction component peaks around
$A_H = 2.1$.  The stronger, main peak is around $A_H = 2.5$ and the
higher extinction plateau region stretches over about 1.5\,mag in
$A_H$ centred around $A_H = 3.5$.  There is also an small, even higher
extinction tail extending between $4.5 \leq A_H \leq 6$.

The $A_{\ks}$ extinction distribution follows approximately the same
distribution as the $A_J$ and $A_H$ above, in the range $0.5 \leq
A_{\ks} \leq 4.5$.  For $A_{\ks}$ the low extinction component is
almost not present, and the higher extinction plateau is more
pronounced.  The main, largest component of the distribution is
centred on $A_{\ks} = 1.1$, and the higher extinction plateau extends
from $1.3 \leq A_{\ks} \leq 2.5$ with two apparent peaks at $A_{\ks} =
1.55$ and $A_{\ks} = 2.05$.  The high extinction tail extends from
$A_{\ks} = 2.5$ out to approximately $A_{\ks} = 4.5$.

\begin{figure*}
  \begin{center}
    \includegraphics[trim= 2mm 0mm 2mm 0mm, clip, width=0.32\textwidth]{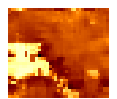}
    \includegraphics[trim= 2mm 0mm 2mm 0mm, clip, width=0.32\textwidth]{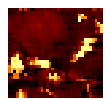}
    \includegraphics[trim= 2mm 0mm 2mm 0mm, clip, width=0.32\textwidth]{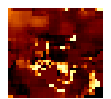}
    \caption[]{Maps of the measured extinction in three of the VLT
      fields in $A_{J}$.  Each map fills the full extent of the
      $2.5\arcmin \times 2.5\arcmin$ VLT fields, and each pixel within
      each map represents the extinction in a region $5\arcsec \times
      5\arcsec$.  The colour scale represents variation in the
      absolute extinction in the range: $0 \leq A_{J} \leq 10$.}
    \label{f:extmaps}
  \end{center}
\end{figure*}

Figure \ref{f:allext} shows the three extinction distributions plotted
on the same graph. This shows the differing values of the extinction
compared to each other, and the similarity of the shapes of the three
distributions.  The three distributions appear to be a single
distribution that is stretched at shorter wavelengths ($A_J$ is
broader and lower than $A_{\ks}$ for example).  Figure \ref{f:extmaps}
shows the extinction maps of three example fields showing the degree
of variation within each $2.5\arcmin \times 2.5\arcmin$ field.  Each
pixel in these images corresponds to one of the grid points separated
by 5\arcsec\ as outlined in Section \ref{s:calcext}, and the colour
scale represents $0 \leq A_H \leq 10$.

\subsection{Comparison with non-varying Extinction Law}
\label{s:nonvary}

In order to test the significance of the variation in the extinction
law, we re-ran our extinction calculation twice, once allowing the
value of the extinction law parameter $\alpha$ to vary from field to
field but not within the field and a second time using the mean
extinction law parameter for all of the fields, $\alpha = 2.64$ (see
Figure \ref{f:alfhisto}). To measure the effects, we compared how the
value of absolute extinction calculated using equations \ref{e:ext}
changes for the different colour excesses (for example: the difference
between $A_J$ calculated from $\left<E(\jh)\right>$ and
$\left<E(\jk)\right>$).  We measured the difference using a
$\chi_{\nu}^2$ test for each value of $A_{\lambda}$.

For the fully varying, V-NICE extinction calculation method that we
are presenting in this paper, the $\chi_{\nu}^2$ test returned values
in the range $1.34 - 2.37$, indicating that the variation between the
measures is within one or two $\sigma$ of the distribution for almost
all cases (Table \ref{t:extcompare}).  Allowing the extinction law to
vary from one field to another, but not within the field, we find that
the differences in the values of absolute extinction calculated from
different colour excesses result in a $\chi_{\nu}^2$ test returning
results of $10.78 - 15.35$. This illustrates that there is a
statistically significant difference between the distributions of
absolute extinction calculated for each wavelength from the two
applicable colour excesses when keeping the extinction law constant
across individual fields rather than allowing it to vary within a
given field.  In other words, if the value of $\alpha$ is held
constant over the 2.5 arcmin$^{2}$ of any one VLT field, different
values of the absolute extinction (i.e.: $A_J$) will be calculated
depending upon which colour excess is used (i.e.:
$\left<E(\jh)\right>$ or $\left<E(\jk)\right>$). This illustrates that
it is important to allow for variation in the extinction law on as
small a scale as is possible for any dataset to derive the most
accurate measure of the extinction in a given field.

The final re-run of the extinction calculation used a single
extinction law, fixed at the average value for all the fields ($\alpha
= 2.64 \pm 0.52$) as measured in Figure \ref{f:alfhisto}, to convert
colour excess to absolute extinction for all 26 of the VLT
fields. This is similar to the method that has previously been used to
calculate the extinction throughout the Galaxy where variations in
extinction law have rarely been taken into account. In this case there
were obvious differences in the calculated values of absolute
extinction from the different colour excesses, revealed by the
$\chi_{\nu}^2$ test which returned values of $19.85 - 55.05$.

\subsection{Intrinsic effects of filters}
\label{s:filters}

We note that the value of the extinction law parameter $\alpha$ that
we measure from our data is significantly different from the more
generally accepted value $\alpha \sim 2$ \citep[][and
  others]{riek85,card89,math90,nish06}.  Here we consider that the
method that we use to calculate the extinction law parameter may be
affecting the measured value.  In equation \ref{e:extlaw}, it is
assumed that the filters in the telescope system used are infinitely
thin, whereas in reality they admit light over a range of wavelengths
with varying efficiency, the transmission profile.  It is possible
that in sampling a larger bandwidth of the stellar spectra than only
at the central wavelengths as assumed in our methods; the filter
transmission profiles are the cause of the difference of our measured
values of $\alpha$ from previous studies.

To explore this, we simulated the effects of using both a narrow
idealized filter, as assumed in our extinction calculation method, and
the real transmission functions of the VLT-ISAAC filters as given on
the ESO website\footnote{ \url{
    http://www.eso.org/sci/facilities/paranal/instruments/isaac/inst/isaac_img.html}}.
To simulate the narrow filters we used thin gaussians with peak
transmission efficiency of 1 and width of $100\,{\rm \AA}$ centred on
the quoted central wavelengths of the VLT-ISAAC filters.  For the
actual filter case, we could only apply the filter transmission
functions as there is no comprehensive data available on the ESO
website which takes into account the additional effects of the
atmosphere, telescope and instrument optical path on light
transmission in VLT-ISAAC.  Experience suggests that these effects
will be predominately attenuation of the light transmitted through the
system and so the filter transmission function will have the largest
effect on variation in the measured extinction.  We used a sample of
stellar spectra, G, K and M type giants from the \citet{pick98}
libraries, and applied extinction to all spectra with a power-law
wavelength dependence $A_{\lambda} \propto \lambda^{-\alpha}$.  We
used $\alpha = 2$ for the extinction applied to the spectra as that is
the most commonly used value for extinction in the \nir\ and it also
allows us to compare the results to the values we measured with our
V-NICE method.  We then used the \texttt{IRAF} package
\texttt{synphot}\footnote{ \url{
    http://www.stsci.edu/resources/software_hardware/stsdas/synphot}}
to pass the extincted standard star spectra through the two filter sets
described above.  We used the magnitudes output by the \texttt{synphot}
package for the different filters to calculate colour excess for the
spectral types, and from the colour excess calculated the extinction
law parameter $\alpha$ using equation \ref{e:extlaw}.

\begin{table}
  \begin{center}
    \caption[$\chi_{\nu}^2$ tests of a single extinction law, a one
      field extinction law and a completely variable extinction
      law]{Results of the tests of the effect of extinction law
      variation on calculated absolute extinction for three
      scenarios:  one with a fully variable extinction law ({\it top}),
      one with an extinction law fixed for individual fields ({\it
      middle}) and one with a single extinction law for all fields
      ({\it bottom}). For each instance, the differences in the values
      of absolute extinction for a single wavelength calculated from
      the two applicable colour excesses are compared with a
      $\chi_{\nu}^2$ test. The results of these $\chi_{\nu}^2$ tests
      are presented against the colour excess error used in their
      calculation. It shows that the fully variable extinction law
      produces the most consistent calculation of absolute extinction
      from different colour excesses and that a fixed extinction law
      produces the worst.}
    \begin{tabular}{lcllc}
      \hline
      \hline
      \multicolumn{1}{c}{Colour excess} & & & \multicolumn{1}{c}{Colour excess} & \\
      \multicolumn{1}{c}{error used} & $\chi_{\nu}^2$ & & \multicolumn{1}{c}{error used} & $\chi_{\nu}^2$ \\
      \hline
       & & & &  \\
      \multicolumn{5}{c}{Fully variable extinction law}\\
      \hline
      $A_J \left<E(\jh)\right>$    & 2.3675 & & $A_J \left<E(\jk)\right>$     & 2.2644 \\
      $A_H \left<E(\jh)\right>$    & 1.3412 & & $A_H \left<E(\hk)\right>$     & 1.8251 \\
      $A_{\ks} \left<E(\jk)\right>$ & 1.7369 & & $A_{\ks} \left<E(\hk)\right>$ & 2.0890 \\
      \hline
       & & & &  \\
      \multicolumn{5}{c}{Single extinction law for each field}\\
      \hline
      $A_J \left<E(\jh)\right>$    & 12.0757 & & $A_J \left<E(\jk)\right>$     & 10.7796 \\
      $A_H \left<E(\jh)\right>$    & 12.7166 & & $A_H \left<E(\hk)\right>$     & 15.3468 \\
      $A_{\ks} \left<E(\jk)\right>$ & 12.4007 & & $A_{\ks} \left<E(\hk)\right>$ & 14.1556 \\
      \hline
       & & & &  \\
      \multicolumn{5}{c}{Extinction law $\alpha = 2.64$ applied to all fields}\\
      \hline
      $A_J \left<E(\jh)\right>$    & 19.8532 & & $A_J \left<E(\jk)\right>$     & 21.2689 \\
      $A_H \left<E(\jh)\right>$    & 27.8859 & & $A_H \left<E(\hk)\right>$     & 55.0459 \\
      $A_{\ks} \left<E(\jk)\right>$ & 34.4246 & & $A_{\ks} \left<E(\hk)\right>$ & 39.5893 \\
      \hline
      \hline
    \end{tabular}
    \label{t:extcompare}
  \end{center}
\end{table}

We found that the case of thin gaussians returns the true value of
$\alpha = 2.00 \pm 0.05$, as expected.  For the actual ISAAC filter
transmission functions, the values of $\alpha$ measured for the
different spectral types of the stars were in the range $1.75 \leq
\alpha \leq 1.85$.  The variation in the measured extinction law
corresponded to variation in the spectral types of the stars, earlier
type stars resulted in a lower measured value of $\alpha$, and later
types resulted in a higher measured value of $\alpha$ (within the
range given above).

To average the effects of the different spectral types, and to
recreate our actual method used to calculate extinction, namely
sampling a variety of stars over a region of the sky, we selected 40--50
random spectral types, adding 5\% noise to the measured magnitudes in
all three bands, and calculated an average value for the extinction
law for the combined colours of the stars.  In 1000 repetitions, we
measured a value of $\alpha = 1.81 \pm 0.11$ (the error being
consistent with the random noise added to the simulated magnitudes).

\section{Discussion}

We have identified two significant sources of potential error in the
calculation of the extinction in the \nir. First, there may be
substantial significant variation in the extinction-wavelength
power-law relationship parameter $\alpha$ (where $A_{\lambda} = C
\times \lambda^{-\alpha}$) and in the absolute value of extinction
$A_{\lambda}$, especially in heavily extinguished regions of the
Galactic Bulge and Plane.  Second, the measured values of $\alpha$ are
influenced by the transmission profile of the instrument used to
obtain the photometric data on which the extinction calculation is
based.

\subsection{Extinction calculation}

We have applied our new V-NICE extinction calculation method which
measures variation on small spatial scales in both the extinction law
and the absolute extinction to fields distributed ``randomly''
throughout the \nb, avoiding known regions of star formation and
structures such as the Arches and other clusters. As such, these
fields offer a deeper, representative sample of the ``general'' \nb\
population than it has been possible to obtain from previous
observations such as \tmass.  As we have demonstrated in
\citet{gosl07}, this method produces self-consistent measures of the
extinction that can recover real photometric values for the extinction
corrected stars.  Until similar analysis can be undertaken on
forthcoming \nir\ observations of the entire \nb, using data from the
UKIDSS and VISTA surveys, this work represents the most generalised
and representative study of the extinction of the overall \nb.

\subsection{Extinction Law}
\label{s:extlawdisc}

In all fields we see variation in the measured extinction law on
scales of $\sim\,5 - 10\arcsec$, which is also the scale of the
granularity measured in \citet{gosl06}.  We measure a broad range of
values of the extinction law parameter, in the range
$0 \leq \alpha \leq 4$. Fitting a gaussian to the extinction law
parameter distribution, we find that the extinction law in the
\nb\ can be described in the most general terms of $\alpha = 2.64$
with a standard deviation of $0.52$ (see Figure \ref{f:alfhisto}).

We tested the variation in the extinction law parameter as measured
with VLT-ISAAC, Table \ref{t:extcompare} shows the effect of the
variation on the calculated absolute extinction when using different
colour excesses. Only in the case of allowing full variation of the
extinction law are the absolute extinction values calculated from
different colour excesses the within $3\sigma$ limits.  For the cases
of a more general extinction law, a single value for each field and a
single value for all fields, the differences introduced by the
calculation of absolute extinction from different colour excesses are
beyond what can be explained as statistical variations in the
distributions (i.e.: differences greater than
$3\sigma$). Additionally, in \citet{gosl07}, we demonstrate that the
use of a fully variant extinction law is the only way to obtain
consistent photometry for stellar sources in our \vlt\ fields and, by
extension, throughout the \nb.

\subsection{Absolute Extinction}
\label{s:extdisc}

We see considerable levels of variation in the measured absolute
extinction on very small scales resulting in overall extinction ranges
in the \nir\ of $4.5 \leq A_J \leq 10$, $1.5 \leq A_H \leq 6.5$ and
$0.5 \leq A_{\ks} \leq 4.5$.  The spread of the measured values of
absolute extinction are similar for all three bands.  Each has a small
amount of very low levels of extinction, a large fraction centred on a
single value, with a ``plateau'' of increasing values of extinction
with a final, very small high extinction tail.  That these
distributions are similar, with the shorter wavelength distribution
``stretched'' indicates that all three \nir\ bands are measuring the
same extincting material distribution.  The variations we measure are
on scales of $\sim\,5 - 10\arcsec$, the same scales at which we
measured variation in the stellar distribution in \citet{gosl06}.  We
interpret the distribution of different values of extinction as
representing the distribution of these dense, obscuring clouds
identified in our previous paper.  The largest peak in all three
distributions, at the lower end of the overall extinction distribution
represents the ``field'' value away from these extincting structures,
the higher extinction ``plateaux'' represent the extinction in the
denser regions, the size scale of which we measured in \citet{gosl06}.
It is thus important to consider both the size of these structures,
and their distribution when calculating the extinction in the \nb\ as
applying a ``field'' value for the extinction could underestimate
extinction by a factor of 2--3 if the target is close to or in one of
these dense, complex regions.

The values of $A_{\ks}$ are we measure similar to the values that have
previously been measured, with the majority of the $A_{\ks}$
distribution in the region of 1--3; however, due to the variable and
generally greater value of the extinction law parameter $\alpha$, the
values of $A_J$ and $A_H$ in this work are considerably higher than
have previously been measured.  Both our calculation of the extinction
from different colour excesses (Section \ref{s:nonvary}), and the
extinction correction of stars in \citet{gosl07} for photometric
identification indicate that this higher value of the extinction law
parameter $\alpha$ and the small scale variations are the correct
method to use in the \nb.  The higher value of $\alpha$ means that
previous works \citep[][and others]{catc90,schu99,lomb01} which quoted
values of $A_V$ based on measures of extinction in the near-infrared
will have significantly underestimated the values of $A_V$.  As an
example, the most commonly used conversion factor from near-infrared
to visual is $A_V/A_{\ks} = 9$ \citep{riek85}; more recently
\citet{nish08} have reported that in areas of the outer Bulge, they
measured a higher conversion factor of $A_V/A_{\ks} \sim 16$.  Using
the conversion factors of \cite{mart90} and \cite{riek85} to derive
the mean relation between $A_{\ks}$ and $A_V$, our work would suggest
a conversion factor of $A_V/A_{\ks} = 28.7^{+17.2}_{-11.7}$ in the
\nb\ for $\alpha = 2.64 \pm 0.52$, 3 times higher than the most
commonly used value.

\subsection{Measuring the Extinction}

We recommend that wherever possible, i.e. if data from at least three
infrared bands is available for a given location, a specific
calculation of the extinction be carried out to obtain accurate
stellar photometry in heavily extinguished regions of the sky such as
the \nb.  Due to the very small scale, real variation that we measure
in the extinction law in this region, it is extremely important that
as local a value as possible is measured.  Using our V-NICE method
allows for the very small spatial variations in the extinction law
parameter $\alpha$ as well as the absolute extinction, producing the
most accurate measure of the extinction at different wavelengths in
the \nir.

\subsection{Filter transmission effects}

As seen in Section \ref{s:filters}, the transmission function of the
filter-set used in observations has an effect on the value of the
extinction law parameter measured using our method.  The results
suggest that only for an idealized very thin filter profile will the
``real'' value of the extinction law be measured using this method.
For the broader filter profiles of real instruments, the spectral
slope of the star being observed will influence the value of the
extinction law measured.  The spatial variations in the extinction law
that we observe are not affected by this property, only the exact
numerical value measured for $\alpha$.  The result of this is that the
extinction law parameter $\alpha$ and the absolute extinction measured
from a set of observations using a unique telescope-instrument setup
can only be applied to data from other observations using the same
telescope-instrument combination without the need to account for the
filter transmission.  If the extinction correction values measured
from one dataset are applied to data from observations obtained using
other telescope-instrument combinations, an error will be induced by
the differing filter transmissions.

\section{Conclusions - A new Extinction measure}

We have studied the extinction in the \nir\ in a sample of
representative regions of the \nb\ based on observations with
VLT-ISAAC.  We have found a high degree of spatial variability in both
the extinction law parameter $\alpha$ as well as the absolute
extinction in these regions.  We have used a new method, a variable
\nir\ colour excess, V-NICE, to measure these variations.  Only when
the value of the extinction law parameter $\alpha$ as well as the
absolute extinction is allowed to vary can self-consistent values for
the extinction at different wavelengths in the \nir\ be calculated.

For our observations, using VLT-ISAAC, the peak of the extinction law
parameter $\alpha$'s distribution for the \nb\ is at a higher value
than previously thought, described by a gaussian with $\mu=2.64$ and
$\sigma=0.52$. We allowed the extinction law to vary when calculating
the absolute extinction at different wavelengths from colour excesses
and show that only when this variation is allowed can self-consistent
levels of extinction be recovered.  The variations of both the
extinction law parameter $\alpha$ and the absolute extinction are seen
on scales as small as $5\arcsec$.  We measure extinction in the
\nir\ of $4.5 \leq A_J \leq 10$, $1.5 \leq A_H \leq 6.5$ and $0.5 \leq
A_{\ks} \leq 4.5$.  The higher overall value of the extinction law
parameter $\alpha$ results in a higher conversion factor between
$A_J$, $A_H$ and $A_{\ks}$.  This suggests that previously derived
levels of $A_{\lambda}$ in the \nb\ from \nir\ extinction measures may
need to be reconsidered.  

We have also found that the extinction law parameter $\alpha$ has a
dependency on the transmission function of the instrument.  We
simulated the effects of the instrument filter transmission profiles
on the measured value of the extinction law parameter $\alpha$ and
showed that in the case of idealized narrow filters, our method
recovers the true value of the extinction law, but for the real ISAAC
filter transmission profiles, the result measured is $\sim\,10\%$
below the true value.

The effects of the small scale variation in the extinction law as well
as the absolute extinction are the dominant source of errors when
correcting for extinction in the \nb.  Furthermore, we suggest that
the potential for small-scale physical variations in the extinction
should be considered when performing photometry in any significantly
extinguished area of the Galactic Bulge or Plane.  The method
presented here is sufficient to obtain accurate stellar photometry for
multi-colour data from any \nir\ telescope-instrument combination, and
can be used to correct for extinction to allow identification of
stellar types by comparison to photometric standards.  The exact
extinction values measured will be relative to the
telescope-instrument combination.  Unless the absolute value of
extinction law is required, or the measured extinction will be used to
correct the colours of stars from datasets from other combinations of
telescope and instrument, the effects of the filter transmission
effects discussed here need not be considered.  However, if the
measured extinction law value is used to reddening-correct data from
different telescope-instrument combinations, the user should be aware
of the errors that will be induced by the different filter
transmission profiles.

Due to these two effects, we suggest that the use of a generic
``universal'' IR extinction law is not appropriate in many cases,
especially when observing regions of the Galaxy known to suffer
significant extinction.  We recommend that whenever possible, a local
extinction calculation should be undertaken as this provides the most
accurate possible photometry for Bulge stars \citep{gosl07}.  We
suggest that the measure of granularity from \citet{gosl06} be used to
determine the scale at which the extinction is variable and this be
used to determine a limit to the area over which stars can be sampled
in calculating the extinction.  The methods we present here are an
effective and accurate way of calculating the extinction for a dataset
to correct the extinction towards sources in the \nb\ in the
near-infrared.

\subsection*{Acknowledgements}

The authors would like to thank the referee Prof. Edward Fitzpatrick
for his extremely thorough reading of this manuscript and detailed
points which have greatly improved this work.  AJG would like to thank
S.T.F.C. and Suomen Akatemia for grants funding his research.  KMB thanks
the Royal Society and the Leverhulme Trust for research support.  This
paper is based on observations made with the ESO VLT at Paranal under
imaging programme ID 071.D-0377(A).

\end{document}